\documentclass[aps,prd,preprint,tightenlines,superscriptaddress,nofootinbib]{revtex4}
\usepackage{amsmath,amssymb}
\usepackage{graphicx,tabularx}
\renewcommand{\textfraction}{0.0}

\def\bq{\begin{equation}}
\def\eq{\end{equation}}
\def\ba{\begin{eqnarray}}
\def\ea{\end{eqnarray}}

\begin{document}

\preprint{$
\begin{array}{l}
\mbox{UB-HET-05-01}\\
\mbox{FERMILAB-PUB-05-542-T}\\
\end{array}
$}

\date{\today}

\title{Improved Measurement of $ttZ$ Couplings at the LHC}

\author{U.~Baur\footnote{baur@ubhex.physics.buffalo.edu}}
\affiliation{Dept. of Physics,
State University of New York, Buffalo, NY 14260, USA}

\author{A.~Juste\footnote{juste@fnal.gov}}
\affiliation{Fermi National Accelerator Laboratory, Batavia, IL 60510,
USA} 

\author{L.H.~Orr\footnote{orr@pas.rochester.edu} and
        D.~Rainwater\footnote{rain@pas.rochester.edu}}
\affiliation{Dept. of Physics and Astronomy, University of Rochester,
Rochester, NY 14627, USA}

\begin{abstract}
\vspace{5mm}
We consider QCD $t\bar{t}Z$ production at the LHC with
$Z\to\bar\nu\nu$ and all-hadronic $t\bar{t}$ decays, {\it i.e.} $pp\to
p\llap/_Tb\bar{b}$+4~jets, as a tool to measure $ttZ$ couplings.  This
channel has a significantly larger cross section than those where the
$Z$ boson decays leptonically.  However, $t\bar{t}$, $b\bar{b}+4$~jet,
$t\bar{t}j$ and $t\bar{t}jj$ production give rise to potentially large
backgrounds.  We show that these processes can be suppressed to an
acceptable level with suitable cuts, and find that adding the
$p\llap/_Tb\bar{b}+4$~jet channel to the final states used in previous
$ttZ$ couplings analyses will improve the sensitivity by $10-60\%$.
We also discuss how the measurement of the $ttZ$ couplings may
constrain Little Higgs models.
\end{abstract}

\maketitle

\newpage

%------------------------------------------------------------------------
%------------------------------------------------------------------------

%------------------------------------------------------------------------
\section{Introduction}
\label{sec:sec1}
%------------------------------------------------------------------------

Although the top quark was discovered more than ten years
ago~\cite{topcdf,topd0}, many of its properties are still only poorly
known~\cite{Chakraborty:2003iw}.  In particular, the couplings of the
top quark to the electroweak (EW) gauge bosons have not yet been
directly measured.~\footnote{Although the measurement of the $W$ boson
helicity in top quark decay~\cite{whelTEV} may be regarded as a
constraint on top quark couplings.}  The large top quark
mass~\cite{PDF2005} suggests that it may play a special role in EW
symmetry breaking (EWSB).  New physics connected with EWSB may thus be
found first in top quark precision observables.  A possible signal for
new physics is a deviation of the any of the $tt\gamma$, $ttZ$ or
$tbW$ couplings from the values predicted by the Standard Model (SM).
For example, in technicolor and other models with a strongly coupled
Higgs sector~\cite{examples}, and in Little Higgs
models~\cite{Berger:2005ht}, anomalous top quark couplings may be
induced at the $5-10\%$ level.

Current data provide only weak constraints on the top quark couplings
to EW gauge bosons, except for the $ttZ$ vector and axial vector
couplings, which are rather tightly but indirectly constrained by LEP/SLC 
$Z$-pole data (see Ref.~\cite{Baur:2004uw} and references therein);
and the right-handed $tbW$ coupling, which is severely bounded by the
observed $b\to s\gamma$ rate~\cite{Larios:1999au}.  Future collider
experiments offer many possibilities to probe the EW top quark
couplings.  The most promising ones with respect to the $ttV$
($V=\gamma,\, Z$) couplings are provided by an $e^+e^-$ linear
collider via $e^+e^-\to\gamma^*/Z^*\to
t\bar{t}$~\cite{Grzadkowski:1998bh,Grzadkowski:2000nx,Lin:2001yq,Abe:2001nq,Aguilar-Saavedra:2001rg,Frey:1995ai,Ladinsky:1992vv},
and the LHC via $t\bar{t}V$
production~\cite{Baur:2004uw,Zhou:1998bh,Baur:2001si}.

At an $e^+e^-$ linear collider operating at $\sqrt{s}=500$~GeV and
with an integrated luminosity of $100-200$~fb$^{-1}$, one could
measure the $ttV$ couplings in top pair production with a few-percent
precision~\cite{Abe:2001nq}.  However, the process
$e^+e^-\to\gamma^*/Z^*\to t\bar{t}$ is sensitive to both $tt\gamma$
and $ttZ$ couplings simultaneously, and significant cancellations
between the various couplings can occur.  At hadron colliders,
$t\bar{t}$ production is so dominated by the QCD processes
$gg,q\bar{q}\to t\bar{t}$ that a measurement of the EW neutral
couplings via $q\bar{q}\to\gamma^*/Z^*\to t\bar{t}$ is hopeless.
Instead, they can be measured in QCD $t\bar{t}Z/\gamma$ production and
radiative top quark decays in $t\bar{t}$ events ($t\bar{t}\to\gamma
W^+W^- b\bar{b}$).  Each of the processes is sensitive to the EW
couplings of only the boson emitted: $Z$ and photon independently.
This obviates having to disentangle potential cancellations between
the different couplings.  In these three processes one can also hope
to separate the dimension-four and -five couplings which appear in the
effective Lagrangian describing $ttV$ interactions.

In Ref.~\cite{Baur:2004uw} we presented a detailed analysis of
$t\bar{t}V$ production at hadron colliders.  We found that while the
$tt\gamma$ couplings can be measured at the LHC with a precision of
typically a few percent, the bounds on the $ttZ$ couplings are a
factor $3-10$ weaker, in particular for the vector and axial vector
couplings, $F_{1V}^Z$ and $F_{1A}^Z$.  A major factor which limits
their sensitivity is the relatively small cross section for
$t\bar{t}Z$ production when one requires the $Z$ boson to decay
leptonically (where we mean $\ell^+\ell^-$ ($\ell=e,\mu$) throughout),
as in our previous analysis.

In this paper, we extend our $t\bar{t}Z$ analysis to the case where
the $t\bar{t}$ pair decays hadronically and the $Z$ boson decays into
neutrinos.  Due to the larger branching ratio for $Z\to\bar\nu\nu$
relative to leptonic decays, this effectively triples the number of
events which can be utilized, thus our present analysis may
significantly improve the limits on anomalous $ttZ$ couplings.
However, the increased statistics comes at the price of a background
which is potentially much larger than the signal.  After reviewing the
definition of the $ttZ$ couplings, we present a detailed discussion of
the signal and all relevant backgrounds (Sec.~\ref{sec:sec2}), showing
that the most important backgrounds can be adequately suppressed by
imposing suitable cuts.  In Sec.~\ref{sec:sec3} we derive sensitivity
bounds for the $b\bar{b}\bar\nu\nu$+$4j$ final state and combine them
with those we obtained previously~\cite{Baur:2004uw}.  We also explore
how a $ttZ$ coupling measurement at the LHC may help constrain
parameters of Little Higgs models.  We summarize our findings in
Sec.~\ref{sec:sec4}.

%------------------------------------------------------------------------
\section{Calculation of Signal and Background}
\label{sec:sec2}
%------------------------------------------------------------------------

%------------------------------------------------------------------------
\subsection{Definition of general $ttZ$ couplings}
\label{sec:sec2a}
%------------------------------------------------------------------------

The most general Lorentz-invariant vertex function describing the
interaction of a $Z$ boson with two top quarks can be written in terms
of ten form factors~\cite{Hollik:1998vz}, which are functions of the
kinematic invariants.  In the low energy limit, these correspond to
couplings which multiply dimension-four or -five operators in an
effective Lagrangian, and may be complex.  If the $Z$ boson couples to
effectively massless fermions, the number of independent form factors
is reduced to eight.  In addition, if both top quarks are on-shell,
the number is further reduced to four.  In this case, the $ttZ$ vertex
can be written in the form
\begin{equation}\label{eq:anomvertex}
\Gamma_\mu^{ttZ}(k^2,\,q,\,\bar{q}) = -ie \left\{
  \gamma_\mu \, \left( F_{1V}^Z(k^2) + \gamma_5F_{1A}^Z(k^2) \right)
+ \frac{\sigma_{\mu\nu}}{2m_t}~(q+\bar{q})^\nu 
   \left( iF_{2V}^Z(k^2) + \gamma_5F_{2A}^Z(k^2) \right)
\right\} \, ,
\end{equation}
where $e$ is the proton charge, $m_t$ is the top quark mass,
$q~(\bar{q})$ is the outgoing top (anti-top) quark four-momentum, and
$k^2=(q+\bar{q})^2$.  The terms $F_{1V}^Z(0)$ and $F_{1A}^Z(0)$ in the
low energy limit are the $ttZ$ vector and axial vector form factors.
The coefficients $F_{2V}^Z(M_Z^2)$ and $F_{2A}^Z(M_Z^2)$, where $M_Z$
is the $Z$ boson mass, are related to the the weak magnetic and
($CP$-violating) weak electric dipole moments, $g_t^Z$ and $d_t^Z$.
At tree level in the SM,
\begin{alignat}{2} \notag
F^{Z,SM}_{1V} & = -\frac{1}{4\sin\theta_W\cos\theta_W} 
  \left( 1 - \frac{8}{3} \, \sin^2\theta_W \right) , & \qquad
F^{Z,SM}_{1A} & = \frac{1}{4\sin\theta_W\cos\theta_W} \, , \\[2.mm] \notag
F^{Z,SM}_{2V} & = 0 \, , & \qquad
F^{Z,SM}_{2A} & = 0 \, , 
\end{alignat}
where $\theta_W$ is the weak mixing angle.  The numerically most
important radiative corrections to the vector and axial vector
couplings can be taken into account by replacing the factor
$(1-8\sin^2\theta_W/3)$ in $F^{Z,SM}_{1V}$ by
$(1-8\sin^2\theta_{eff}^t/3)$, where $\sin^2\theta_{eff}^t$ is the
effective mixing angle; and by expressing the remaining factors of
$\sin\theta_W$ and $\cos\theta_W$ in $F^{Z,SM}_{1V,A}$ in terms of the
physical $W$ and $Z$ masses.  Numerically, the one-loop corrections to
$F^Z_{1V,A}$ are typically of ${\cal O}(10^{-2} -
10^{-3})$~\cite{Hollik:1988ii}.  The weak magnetic dipole form factor
$F^Z_{2V}$ receives contributions of the same
magnitude~\cite{Bernabeu:1995gs} at the one-loop level in the SM.
However, there is no such contribution to the weak electric dipole
form factors, $F^Z_{2A}$~\cite{Hollik:1998vz}.

$S$-matrix unitarity restricts $\Delta F^Z_{1V,A}(0)=
F^Z_{1V,A}(0)-F^{Z,SM}_{1V,A}$ to be $\lesssim {\cal O}(1)$ if the
scale of new physics is of the order of a few
TeV~\cite{Baur:2004uw,Hosch:1996wu}.

%------------------------------------------------------------------------
\subsection{Signal}
\label{sec:sec2b}
%------------------------------------------------------------------------

The process $pp\to t\bar{t}Z$ with leptonic $Z$ boson decay was
considered in detail in Ref.~\cite{Baur:2004uw}.  Here we concentrate
on final states where $Z\to\bar\nu\nu$.  Since the $t\bar{t}Z$ cross
section is too small to be observable at the Tevatron, we concentrate
our efforts on the LHC.  The neutrinos escape undetected and, thus,
manifest themselves in the form of missing transverse momentum,
$p\llap/_T$.  If one or both $W$ bosons originating from the top
decay, $t\to Wb$, decay leptonically, the observed states, $\ell
p\llap/_Tb\bar{b}jj$ and $\ell\ell' p\llap/_Tb\bar{b}$, are identical
to those resulting from ordinary $t\bar{t}$ production.  As the
$t\bar{t}$ cross section is more than a factor 1000 larger than that
of $t\bar{t}Z$, it will be very difficult to sufficiently suppress
this background.  We therefore consider only the case where both $W$
bosons decay hadronically,
\begin{equation}\label{eq:react}
pp\to p\llap/_Tb\bar{b}+4j \, .
\end{equation}
We assume that both $b$ quarks are tagged with a combined efficiency
of $\epsilon_b^2=0.4$.  Note that, since there is essentially no phase
space for $t\to WZb$ decays ($BR(t\to WZb)\approx 3\cdot
10^{-6}$~\cite{Mahlon:1994us,Altarelli:2000nt}), $t\bar{t}$ production
with one top decaying into $WZb$ does not contribute to the final
state of Eq.~(\ref{eq:react}).

We perform our calculation for general $ttZ$ couplings of the form of
Eq.~(\ref{eq:anomvertex}).  At LHC energies, $Z$ boson transverse
momenta of at most a few hundred GeV are accessible in $t\bar{t}Z$
production.  The scale of new physics responsible for anomalous $ttZ$
couplings would be expected to be of ${\cal O}(1$~TeV) or higher.
Form factor effects will thus be small and are therefore neglected in
the following.  We also assume that all $ttZ$ couplings are real.  We
otherwise assume the SM to be valid.  Our calculation includes top
quark and $Z$ boson decays with full spin correlations and finite
width effects.  All top quark resonant Feynman diagrams are included.
To ensure gauge invariance of the SM result, we use the so-called
overall-factor scheme of Ref.~\cite{Baur:1991pp} as implemented in the
{\sc madgraph}~\cite{Stelzer:1994ta}-derived code of
Ref.~\cite{Maltoni:2002jr}.

All signal and background cross sections in this paper are computed
using CTEQ6L1~\cite{Pumplin:2002vw} parton distribution functions with
the strong coupling constant evaluated at leading order and
$\alpha_s(M_Z^2)=0.130$.  We set the top quark mass to
$m_t=178$~GeV~\cite{topmass}.\footnote{A recent update of the Tevatron
top mass analysis using Run~II data has resulted in a slightly lower
value, $m_t=172.7$~GeV~\cite{newmt}.  This value leads to a marginally
higher $t\bar{t}Z$ cross section.}  All signal cross sections in this
paper are calculated for factorization and renormalization scales
equal to $m_t$.

The basic acceptance cuts for $p\llap/_Tb\bar{b}$+$4j$ events at the
LHC are
\begin{eqnarray}\label{eq:cuts1}
\nonumber &
p_T(b) > 20~{\rm GeV} \; , \qquad
|\eta(b)| < 2.5       \; , \qquad
\Delta R(b,b) > 0.4   \; , \\
 &
p_T(j) > 30~{\rm GeV} \; , \qquad
|\eta(j)| < 2.5       \; , \qquad
\Delta R(j,j) > 0.4   \; , \qquad
\Delta R(j,b) > 0.4   \; , 
\end{eqnarray}
where $\Delta R=[(\Delta\phi)^2+(\Delta\eta)^2]^{1/2}$ is the
separation in pseudorapidity--azimuth space.  We include minimal
detector effects via Gaussian smearing of parton momenta according to
CMS~\cite{cms} expectations, and take into account the $b$ jet energy
loss via a parameterized function.  To ensure that the LHC experiments
can trigger on the events of interest, we require at least three ($b$-
or non-$b$) jets to have
\begin{equation}\label{eq:cuts2}
p_T(j)>50~{\rm GeV}
\end{equation}
and 
\begin{equation}\label{eq:cuts3}
p\llap/_T>5~{\rm GeV}^{1/2}\sqrt{\sum p_T} \, ,
\end{equation}
where the sum extends over jets and $b$ quarks in the final state.
Furthermore, to reduce the background from non-resonant $Zb\bar{b}+4j$
production and singly-resonant processes such as $pp\to t\bar{b}Zjj$,
we require that the two $b$ quarks and four jets are consistent with
originating from a $t\bar{t}$ pair. This is accomplished by selecting
events which satisfy
\begin{equation}\label{eq:cuts4}
\chi^2_{min}=\min_{b_1j_1j_2b_2j_3j_4~perm}
\left[\chi^2(b_1j_1j_2;b_2j_3j_4)\right] < 3
\end{equation}
where $\chi^2_{min}$ is the minimum of the
$\chi^2(b_1j_1j_2;b_2j_3j_4)$ values of all possible combinations of
jet pairs and $bjj$ combinations, and
\begin{eqnarray}
\chi^2(b_1j_1j_2;b_2j_3j_4)& = &\frac{(m(j_1j_2)-M_W)^2}{\sigma_W^2}
+ \frac{(m(j_3j_4)-M_W)^2}{\sigma_W^2} +
\\ \nonumber
& &   \frac{(m(b_1j_1j_2)-m_t)^2}{\sigma_t^2}
    + \frac{(m(b_2j_3j_4)-m_t)^2}{\sigma_t^2} \; .
\end{eqnarray}
For the $W\to jj$ and $t\to bjj$ invariant mass resolutions we take
$\sigma_W=7.8$~GeV and $\sigma_t=13.4$~GeV~\cite{Beneke:2000hk}.

As we discuss in more detail below, potentially large backgrounds
arise from ordinary $t\bar{t}$ and $b\bar{b}$+$4j$ production where
the four momentum vector of one or more jets is badly mismeasured.  In
contrast to signal events, the azimuthal opening angle
$\Delta\phi(p\llap/_T,p_T(b\bar{b}))$ between the missing transverse
momentum and the transverse momentum of the two $b$ quarks,
\begin{equation}
{\mathbf p}_T(b\bar{b})={\mathbf p}_T(b)+{\mathbf p}_T(b)
\end{equation}
in $t\bar{t}$ and $b\bar{b}+4j$ background events is typically smaller
than $90^\circ$.  The same is also true for the azimuthal opening
angle $\Delta\phi(p\llap/_T,p_T(had))$ between the missing transverse
momentum and the transverse momentum of the four leading non-$b$ jets,
\begin{equation}
{\mathbf p}_T(had)=\sum_{i=1}^4 {\mathbf p}_T(j_i).
\end{equation}
In addition to the cuts listed in
Eqs.~(\ref{eq:cuts1})--~(\ref{eq:cuts4}), we therefore require
\begin{equation}
\label{eq:cuts5}
\Delta\phi(p\llap/_T,p_T(b\bar{b}))>100^\circ, \qquad 
\Delta\phi(p\llap/_T,p_T(had)) >100^\circ~.
\end{equation}
Imposing the cuts listed in Eqs.~(\ref{eq:cuts1})--~(\ref{eq:cuts5}),
and before taking into account $b$-tagging efficiencies, we obtain a
signal cross section of about 3.4~fb.

%------------------------------------------------------------------------
\subsection{Background processes}
\label{sec:sec2c}
%------------------------------------------------------------------------

The potentially most dangerous irreducible background to
$p\llap/_Tb\bar{b}$+$4j$ production originates from $t\bar{t}j$
production, where one top quark decays hadronically, $t\to Wb\to bjj$,
and the other via $t\to Wb\to\tau\nu_\tau b$ with the $\tau$-lepton
decaying hadronically, $\tau\to h\nu_\tau$.  We calculate this process
using tree level matrix elements which include all decay correlations.
Because of its small mass and typically high transverse momentum, we
simulate $\tau$ decays in the collinear approximation.  All $\tau$
decays are calculated following the approach described in
Ref.~\cite{wbf_ll}.  For the probability that a $\tau$-jet is
misidentified as a light quark/gluon jet we assume a constant value of
$P_{h\to j}=20\%$.  Ref.~\cite{tauveto} found that $P_{h\to j}$
decreases with increasing jet transverse momentum, from $20\%$ for
$p_T(j)=20$~GeV to $5\%$ at $p_T(j)=60$~GeV.  Our results for the
$t\bar{t}j$ background thus should be regarded as conservative.

Other contributions to the irreducible background arise from
singly-resonant top quark production ($t\bar{b}Zjj$ and
$\bar{t}bZjj$), and from non-resonant $WZb\bar{b}jj$ and
$Zb\bar{b}+4j$ production.  The calculation of these backgrounds was
discussed in detail in Ref.~\cite{Baur:2004uw} for $Z\to\ell^+\ell^-$
decays.  It is straightforward to adapt the calculation to
$Z\to\bar\nu\nu$ decays.  Finally, $t\bar{t}jj$ production with
$t\bar{t}\to\tau^+\nu_\tau\tau^-\bar\nu_\tau b\bar{b}$ and both
$\tau$-leptons decaying hadronically has to be considered.  We
calculate this background using {\sc alpgen}~\cite{Mangano:2002ea},
treating $\tau$ decays the same as for the $t\bar{t}j$ background
discussed above.

There are also several reducible backgrounds which result from missing
transverse momentum arising from badly-mismeasured jet momenta, or
from not detecting a charged lepton.  The main contributions to the
first category arise from $t\bar{t}$, $b\bar{b}$+$4j$ and $6j$
production.  We calculate the latter two processes using {\sc alpgen}.
QCD $6j$ production contributes only if two light jets are
misidentified as $b$ jets.  To estimate the contribution of this
process we assume the probability for a light jet to be misidentified
as a $b$ jet to be $P_{j\to b}=1/100$.  QCD $t\bar{t}jj$ production
contributes to the background if $t\bar{t}\to\ell^\pm\nu_\ell
b\bar{b}jj$ ($\ell=e,\,\mu$) and the charged lepton is missed.
Likewise, $pp\to t\bar{t}W$ contributes if the $t\bar{t}$ system
decays hadronically, $t\bar{t}\to b\bar{b}$+$4j$, and the lepton in
$W\to\ell\nu$ is missed; or if $t\bar{t}\to\ell^\pm\nu_\ell
b\bar{b}jj$ and $W\to jj$.  We assume that an electron or muon is
missed if $|\eta(\ell)|>2.5$, $p_T(\ell)<10$~GeV, or if either $\Delta
R(\ell,j)<0.4$ or $\Delta R(\ell,b)<0.4$.  Since electrons can be
detected, albeit with reduced efficiency, at rapidities larger than
2.5 using the forward electromagnetic calorimeter, our estimates of
the $t\bar{t}jj$ and $t\bar{t}W$ backgrounds are conservative.  We
calculate the $t\bar{t}jj$ background again using {\sc alpgen}.

Our calculation of the $t\bar{t}j(j)$ backgrounds does not include
contributions from $t\bar{t}(j)$ production where one or both top
quarks decay radiatively, $t\to Wbj(j)$.  Due to the $p\llap/_T$ and
$\chi^2$ cuts of Eqs.~(\ref{eq:cuts3}) and~(\ref{eq:cuts4}), such
contributions are strongly suppressed.

\begin{figure}[t!]
\begin{center}
\includegraphics[width=15cm]{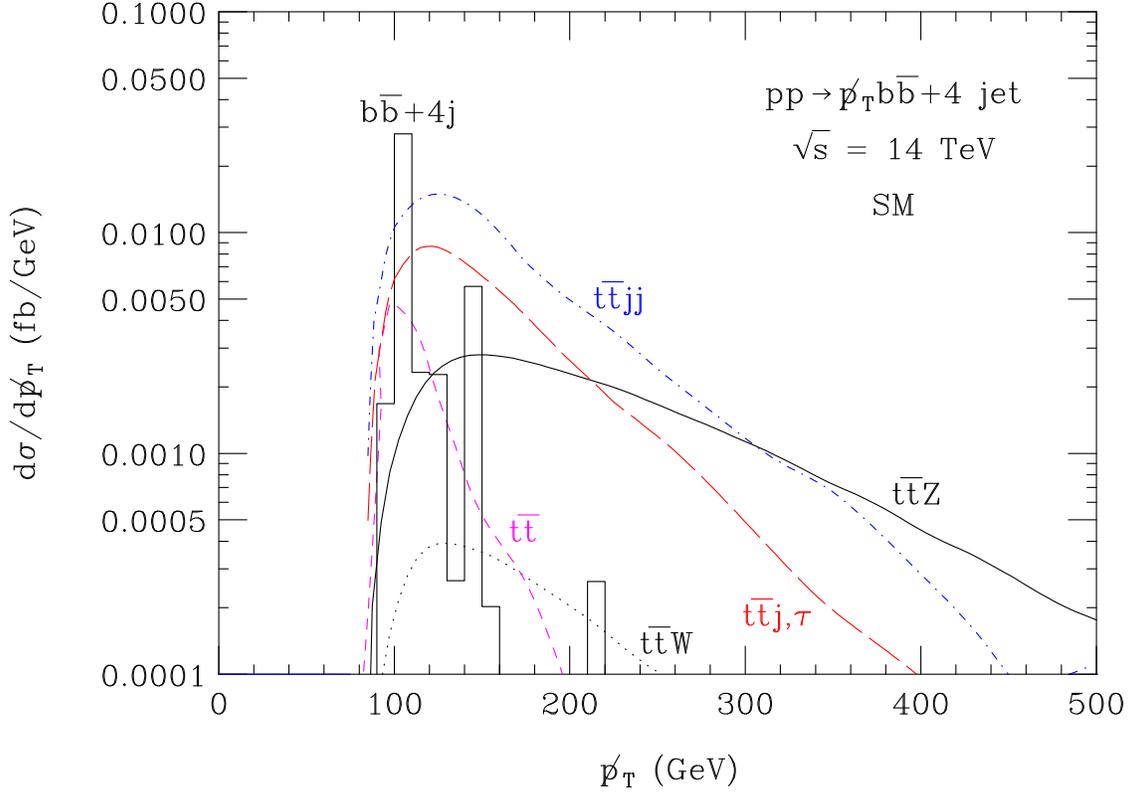} 
\caption[]{The differential cross sections as a function of missing
transverse momentum for $p\llap/_Tb\bar{b}+4j$ production at the LHC.
Shown are the SM predictions for $t\bar{t}Z$ production and various
backgrounds.  We impose the cuts of
Eqs.~(\ref{eq:cuts1}--\ref{eq:cuts5}), but do not include the double
$b$-tag efficiency common to all curves.}
\label{fig:fig1}
\vspace{-7mm}
\end{center}
\end{figure}
In Fig.~\ref{fig:fig1} we show the missing transverse momentum
distributions for the SM $t\bar{t}Z$ signal (solid curve) and for
various backgrounds.  The $p\llap/_T$ requirement of
Eq.~(\ref{eq:cuts3}) implies that $p\llap/_T>80$~GeV.  The most
important backgrounds originate from $t\bar{t}jj$ and $t\bar{t}j$
production.  However, the missing transverse momentum distribution
from these processes falls considerably faster than that of the
signal: for $p\llap/_T>300$~GeV, the SM signal dominates.  The
$t\bar{t}W$ background (dotted line) is about one order of magnitude
smaller than the signal.  The $t\bar{t}$ (dashed line) and
$b\bar{b}$+$4j$ (histogram) backgrounds are important only at low
values of $p\llap/_T$.  Since the missing transverse momentum in these
events originates entirely from $b$-decays and mismeasurements, their
$p\llap/_T$ distributions fall very rapidly.  This is even more the
case for the $6j$ background which we found to essentially vanish
after cuts.  The $t\bar{t}jj$,
$t\bar{t}\to\tau^+\nu_\tau\tau^-\bar\nu_\tau b\bar{b}\to p\llap/_T
b\bar{b}jj$ background is found to be about four orders of magnitude
smaller than the signal, thus is not shown.

The background contributions of $Zb\bar{b}$+$4j$, calculated with {\sc
alpgen}~\cite{Mangano:2002ea}, and $WZb\bar{b}jj$, $t\bar{b}Zjj$ and
$\bar{t}bZjj$ production, calculated with {\sc
madevent}~\cite{Maltoni:2002qb}, are not shown in Fig.~\ref{fig:fig1}.
As discussed in Ref.~\cite{Baur:2004uw}, these cross sections are one
to two orders of magnitude smaller than the signal and can safely be
neglected here.

It should be noted that the background cross sections as calculated at
tree level depend significantly on the choice of factorization and
renormalization scales, $\mu_F$ and $\mu_R$, which were taken to be
$\mu_F=\mu_R=m_t$ in all cases, even for backgrounds without resonant
top quarks.  Including next-to-leading order (NLO) corrections in most
cases significantly reduces the scale dependence of a process.
Unfortunately, the NLO QCD corrections are not presently known for any
of the background processes, except for $t\bar{t}$
production~\cite{Beneke:2000hk,lhctop}.  However, as we shall discuss
in Sec.~\ref{sec:sec3a}, it may be possible to extract the background
cross sections using data.  For the dominant $t\bar{t}j(j)j$
backgrounds this should provide a more accurate estimate of the cross
section than the leading order QCD predictions.

\begin{figure}[t!]
\begin{center}
\includegraphics[width=15cm]{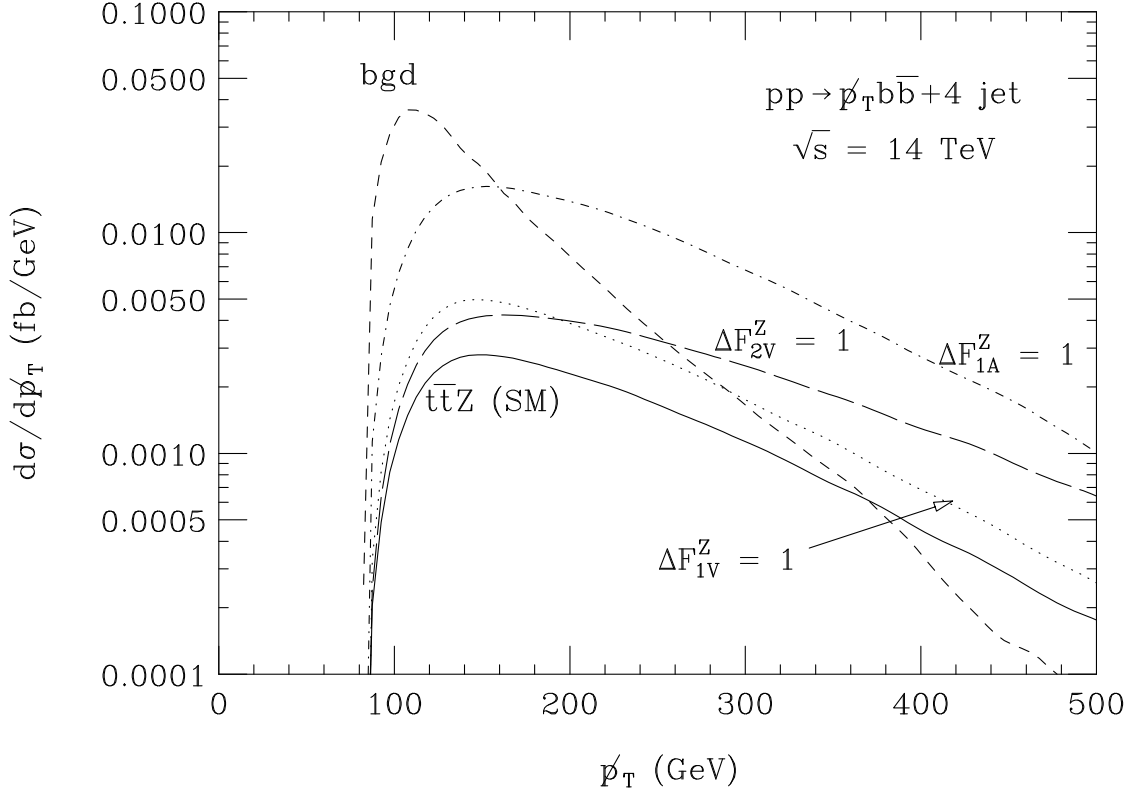} 
\caption[]{The differential cross sections as a function of missing
transverse momentum for $p\llap/_Tb\bar{b}$+$4j$ production at the
LHC.  Shown are the SM predictions for $t\bar{t}Z$ production (solid
line), the combined $t\bar{t}$, $b\bar{b}+4j$, $t\bar{t}j$ and
$t\bar{t}jj$ background, and the predictions for several non-standard
$ttZ$ couplings.  Only one coupling at a time is allowed to deviate
from its SM value.  We impose the cuts of
Eqs.~(\ref{eq:cuts1}--\ref{eq:cuts5}), but do not include the double
$b$-tag efficiency common to all curves.}
\label{fig:fig2}
\vspace{-7mm}
\end{center}
\end{figure}

The $p\llap/_T$ distribution for $pp\to t\bar{t}Z$ in the SM (after
decays), and for various anomalous $ttZ$ couplings, together with the
combined $p\llap/_T$ distribution of the $t\bar{t}$, $b\bar{b}+4j$,
$t\bar{t}j$ and $t\bar{t}jj$ backgrounds, are shown in
Fig.~\ref{fig:fig2}.
Only one coupling at a time is allowed to deviate from its SM
prediction.  Fig.~\ref{fig:fig2} shows that, as in the case when the
$Z$ boson decays leptonically~\cite{Baur:2004uw}, varying $F^Z_{1V,A}$
leads mostly to a normalization change of the SM signal cross section,
hardly affecting the shape of the $p\llap/_T$ distribution.  Thus, the
low-$p\llap/_T$ region contributes most of the statistical weight when
extracting bounds on $F^Z_{1V,A}$.  Since the signal $p\llap/_T$
distribution is approximately proportional to
$(F^Z_{1V})^2+(F^Z_{1A})^2$, it is difficult to disentangle vector and
axial vector couplings in the $p\llap/_Tb\bar{b}$+$4j$ final
state.\footnote{In $t\bar{t}Z$ final states with $Z\to\ell^+\ell^-$,
other distributions, such as the azimuthal (transverse plane) opening
angle between the charged leptons, may be used to help discriminate
between $F^Z_{1V}$ and $F^Z_{1A}$~\cite{Baur:2004uw}.}  Furthermore,
$F^{Z}_{1V,A}=F^{Z,SM}_{1V,A}$ and $F^{Z}_{1V,A}=-F^{Z,SM}_{1V,A}$
yield approximately the same cross sections.

The dimension five couplings, $F^Z_{2V,A}$, on the other hand, lead to
a missing transverse momentum distribution significantly harder than
that predicted by the SM.  As a result, most of the sensitivity to
$F^Z_{2V,A}$ originates from the high-$p\llap/_T$ region.  While the
background is about one order of magnitude larger than the signal
close to $p\llap/_T$ threshold, it is smaller than the signal for
$p\llap/_T>380$~GeV.  As a result, the limits extracted for
$F^Z_{2V,A}$ depend considerably less on the background than those for
$F^Z_{1V,A}$.

As stated before, we require that both $b$ quarks be tagged.  A looser
requirement of at least one tagged $b$ quark would result in a signal
cross section increase of a factor $(2/\epsilon_b-1)$, about 2.2 using
our $b$-tagging assumption.  But this larger signal rate comes at the
expense of an increased background.  In particular, the $6j$ and
$b\bar{b}$+$4j$ backgrounds increase drastically due to the larger
combinatorial background from grouping jets and the tagged $b$ quark.
Detailed calculations would be needed for a quantitative estimate of
the increase.  However, since the single-$b$-tagged background is
probably considerably larger than that for the double $b$-tagged
channel, we do not consider it here.

%------------------------------------------------------------------------
\section{Anomalous Couplings Limits and Model Implications}
\label{sec:sec3}
%------------------------------------------------------------------------

%------------------------------------------------------------------------
\subsection{Limits on anomalous $\boldsymbol{ttZ}$ couplings}
\label{sec:sec3a}
%------------------------------------------------------------------------

The shape and normalization changes of the $p\llap/_T$ spectrum can be
used to derive quantitative sensitivity bounds on anomalous $ttZ$
couplings.  We do this by performing a log-likelihood test on the
distribution and calculating $68\%$ confidence level (CL) limits.  To
calculate the statistical significance, we split the $p\llap/_T$
distribution into bins, each with typically more than five events.  We
impose the cuts described in Sec.~\ref{sec:sec2b} and assume a double
$b$-tag efficiency of $\epsilon_b^2=0.4$.  Except for the $ttZ$
couplings we assume the SM to be valid: the $tbW$ and $ttg$ couplings
can be independently and precisely measured at the LHC in single
top~\cite{Boos:1999dd} and $t\bar{t}$ production~\cite{Beneke:2000hk},
respectively.  We perform multi-parameter fits for the different
anomalous couplings, which we assume to be real.

The log-likelihood function we use to compute confidence levels is
\begin{eqnarray}
\nonumber
-2\log L & = &
-2\left[\sum_i\left(-f_SS_i-f_BB_i+n_{0i}\log(f_SS_i 
                    +f_BB_i) - \log(n_{0i}!)\right)\right]
\\ 
& & + \frac{(f_S-1)^2}{(\Delta f_S)^2}
    + \frac{(f_B-1)^2}{(\Delta f_B)^2} \, ,
\end{eqnarray}
where the sum extends over the number of bins, $S_i$ and $B_i$ are the
number of signal and background events in the $i$th bin, and $n_{0i}$
is the number of SM events in the $i$th bin.  The uncertainties on the
signal and background normalizations are taken into account via two
multiplicative factors, $f_S$ and $f_B$, which are allowed to vary but
are constrained within the relative uncertainties of the signal and
background cross sections, $\Delta f_S$ and $\Delta f_B$,
respectively.

To derive sensitivity bounds, we take into account the dominant
$t\bar{t}jj$, $t\bar{t}j$, $t\bar{t}$ and $b\bar{b}$+$4j$ backgrounds.
We calculate limits for 300~fb$^{-1}$ and 3000~fb$^{-1}$.  An integrated
luminosity of 300~fb$^{-1}$ 
corresponds to 3~years of running at the LHC design luminosity of
${\cal L}=10^{34}\,{\rm cm}^{-2}\,s^{-1}$, while the larger value of
3000~fb$^{-1}$ can be achieved in about 3~years of running at the
luminosity-upgraded LHC (SLHC)~\cite{Gianotti:2002xx}.

As mentioned in Sec.~\ref{sec:sec2}, the dimension five couplings
$F^Z_{2V,A}$ lead to a considerably harder $p\llap/_T$ distribution
than that predicted by the SM.  Most of the sensitivity to these
couplings thus originates from the high missing transverse momentum
region.  In this region, the signal to background ratio is of ${\cal
O}(1)$ or better (cf. Fig.~\ref{fig:fig2}).  The sensitivity bounds on
$F^Z_{2V,A}$ should therefore depend very little on the normalization
uncertainties of signal and background.

For $F^Z_{1V,A}$, however, the situation is different.  Since the
vector and axial vector couplings essentially change only the overall
normalization of the $t\bar{t}Z$ cross section, precise knowledge of
the SM signal cross section is very important.  Most of the
sensitivity to $F^Z_{1V,A}$ comes from the region of small $p\llap/_T$
where the background dominates over the signal.  The achievable
bounds on these couplings are thus expected to depend sensitively on
the signal and background cross section uncertainties.

As explained before, except for the $t\bar{t}$ background, QCD
corrections for neither the signal nor the background cross sections
are known.  The cross sections of the main backgrounds, $t\bar{t}jj$
and $b\bar{b}$+$4j$ production, are proportional to $\alpha_s^4$ and
$\alpha_s^6$, respectively, whereas the signal cross section scales as
$\alpha_s^2$.  The background thus depends more strongly on the
factorization and renormalization scales than the signal.  Its
normalization can be fixed by relaxing the selection cuts
(Eqs.~(\ref{eq:cuts4}) and~(\ref{eq:cuts5})), measuring the cross
section in a background-dominated region of phase space, and then
extrapolating to the analysis region.  Since the cross sections for
$t\bar{t}jj$ and $b\bar{b}$+$4j$ production are large, this should
make it possible to determine the background with an uncertainty
($\Delta f_B$) of a few percent, provided that the systematic
uncertainties can be adequately controlled and that the QCD
corrections do not significantly change the shape of the $p\llap/_T$
distribution of the background.  Exactly how well this will work in
practice remains an open question.

To reduce the signal cross section uncertainty, the NLO QCD
corrections to $t\bar{t}Z$ production must be calculated.  This
appears to be feasible with current techniques.  Once the corrections
are known, the remaining uncertainty ($\Delta f_S$) is likely to be of
order~$10\%$.

To derive quantitative sensitivity limits for anomalous $ttZ$
couplings, we assume $\Delta f_S=0.1$ and $\Delta f_B=0.05$.
Unfortunately, the minimization of $-2\log L$ with respect to $f_S$
and $f_B$ cannot be performed analytically.  Doing it numerically
becomes very time consuming when more than one of the $ttZ$ couplings
are varied at the same time.  However, if one does not allow the
signal and background normalizations to vary independently, {\it i.e.} 
if $f_S=f_B=f$, $\log L$ can be minimized analytically.  In this case,
one finds the minimum of $\log L$ to occur at
\begin{equation}
f=\frac{1}{2}\left(
1-(\Delta f)^2N+\sqrt{(1-(\Delta f)^2N)^2+4(\Delta f)^2N_0}
\right) \, ,
\end{equation}
where
\begin{equation}
N=\sum_i(S_i+B_i)
\end{equation}
is the total number of events, 
\begin{equation}
N_0=\sum_i n_{0i}
\end{equation}
the total number of SM events, and $\Delta f$ is the SM cross section
uncertainty.  The computer time required to derive sensitivity limits
for anomalous $ttZ$ couplings is now much reduced.

The $68\%$ CL bounds on $\Delta F^Z_{1V,A}$ we obtain for $\Delta
f_S=0.1$ and $\Delta f_B=0.05$ agree with those derived for
$f_S=f_B=f$ and $\Delta f=0.3$ within $10-30\%$.  In view of the
current signal and background cross section uncertainties, an analysis
with $f_S=f_B=f$ and $\Delta f=0.3$ when all couplings are varied
should be sufficient.  At first, it may be somewhat surprising that
one needs $\Delta f$ to be significantly larger than $\Delta f_S$ and
$\Delta f_B$ in order to arrive at similar bounds for non-standard
$ttZ$ couplings.  However, varying the signal and background cross
section normalizations separately allows for changes in both the
normalization and the shape of the SM cross section.  On the other
hand, requiring $f_S=f_B=f$ allows for only a change in the
normalization.  Since the anomalous $ttZ$ coupling sensitivity results
from both normalization and $p\llap/_T$ distribution shape changes,
taking into account the uncertainty on both naturally has a relatively
larger impact than if only the normalization uncertainty is included
in the analysis.  One can partially compensate for this effect by
increasing $\Delta f$.

\renewcommand{\bottomfraction}{0.9}
\renewcommand{\textfraction}{0}
\begin{figure}[t!]
\begin{center}
\begin{tabular}{lr}
\includegraphics[width=8.1cm]{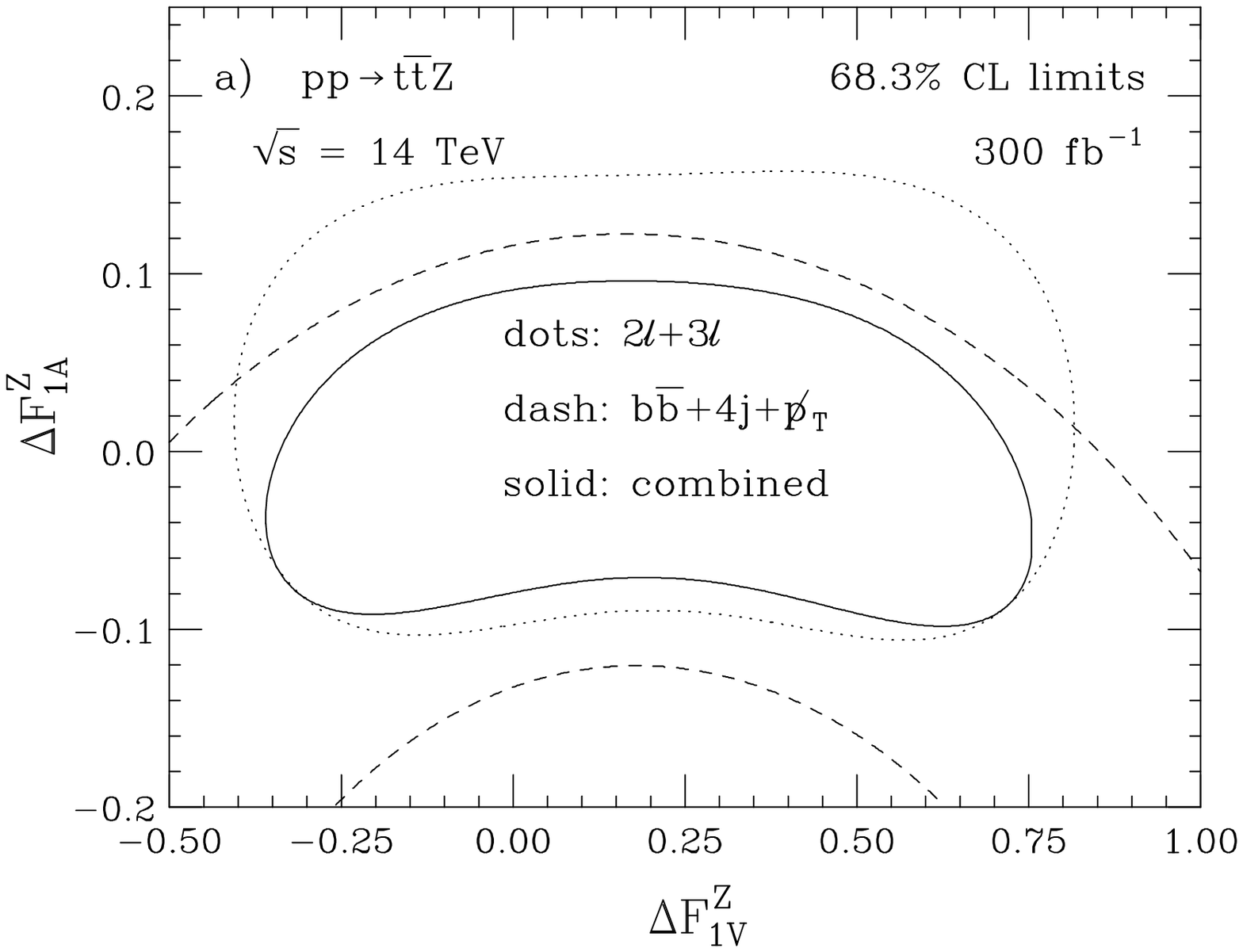} &
\includegraphics[width=8.1cm]{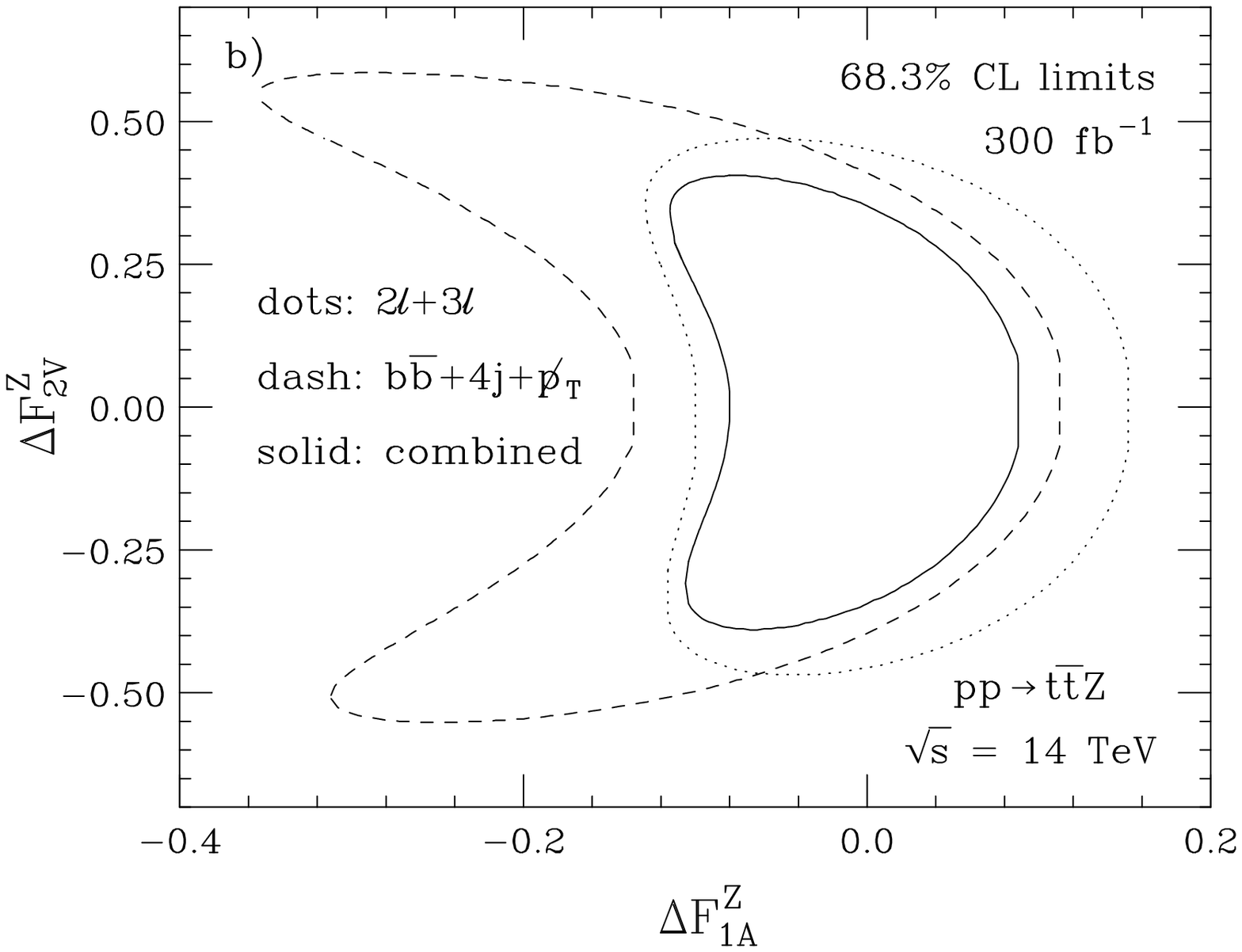} \\
\includegraphics[width=8.1cm]{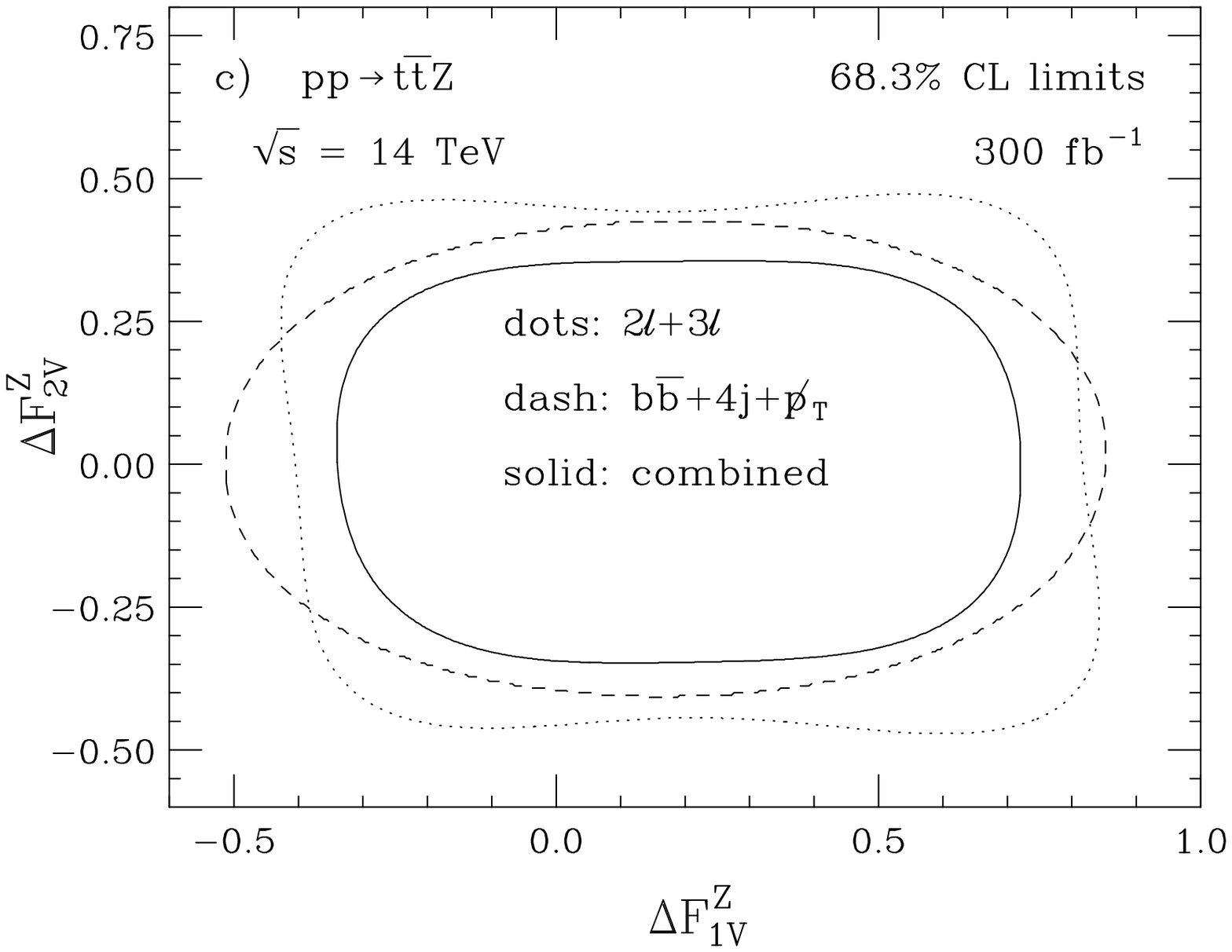} &
\includegraphics[width=8.1cm]{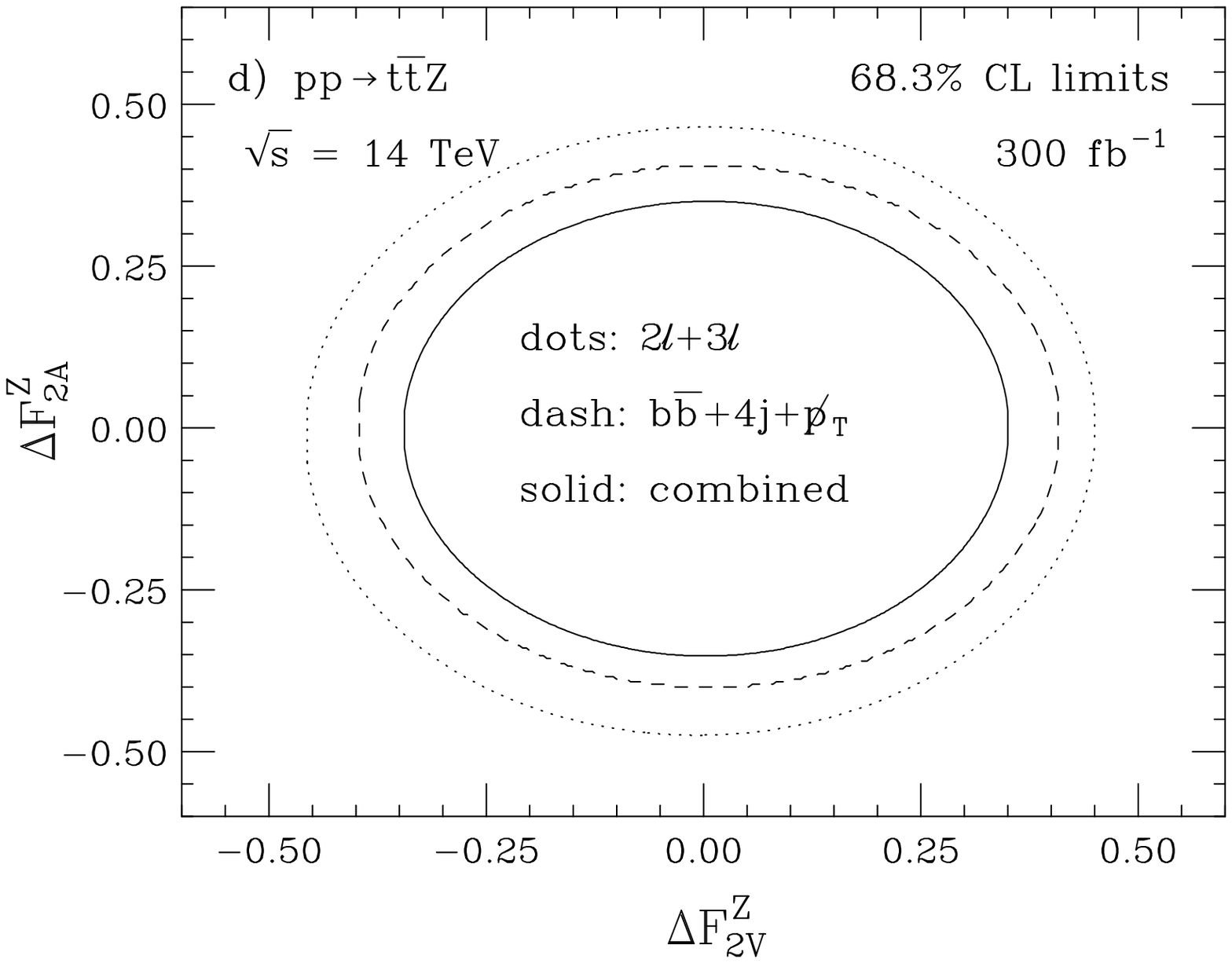} 
\end{tabular}
\caption[]{Projected $68.3\%$ CL bounds on anomalous $ttZ$ couplings
from the LHC with an integrated luminosity of 300~fb$^{-1}$, for: (a)
$\Delta F^Z_{1A}$ versus $\Delta F^Z_{1V}$, (b) $\Delta F^Z_{2V}$
versus $\Delta F^Z_{1A}$, (c) $\Delta F^Z_{2V}$ versus $\Delta
F^Z_{1V}$, and (d) $\Delta F^Z_{2A}$ versus $\Delta F^Z_{2V}$.  Shown
are the limits obtained from the $p\llap/_T b\bar{b}$+$4j$ (dashed)
and the dilepton and trilepton final states (dotted), and the combined
limits (solid).  To derive limits for the dilepton and trilepton final
states, we use the results of Ref.~\protect\cite{Baur:2004uw}.  In
each graph, only those couplings which are plotted against each other
are assumed to be different from their SM values.}
\label{fig:fig3}
\vspace{-7mm}  
\end{center}
\end{figure}
\renewcommand{\bottomfraction}{0.9}
\renewcommand{\textfraction}{0}
\begin{figure}[t!]
\begin{center}
\begin{tabular}{lr}
\includegraphics[width=8.1cm]{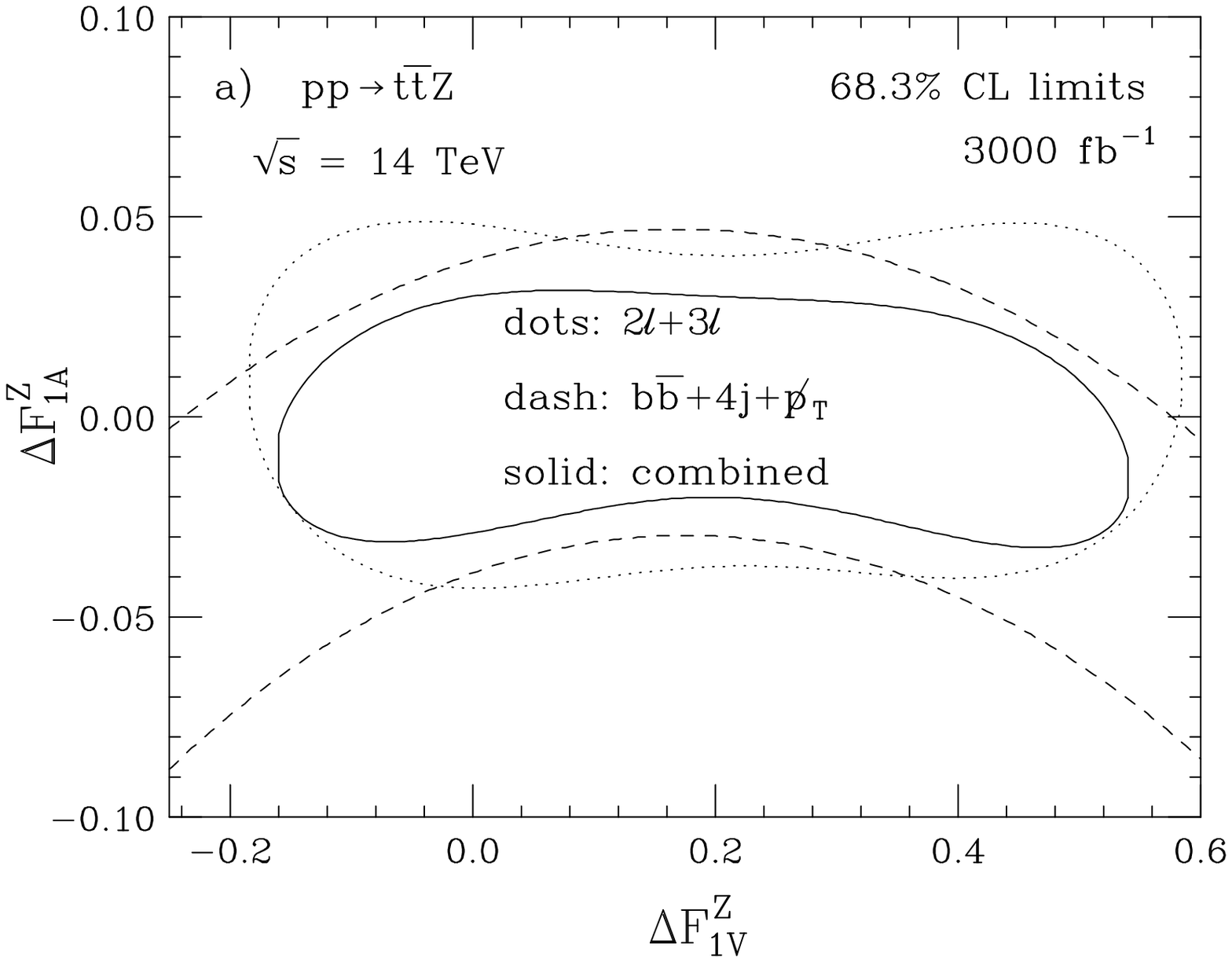} &
\includegraphics[width=8.1cm]{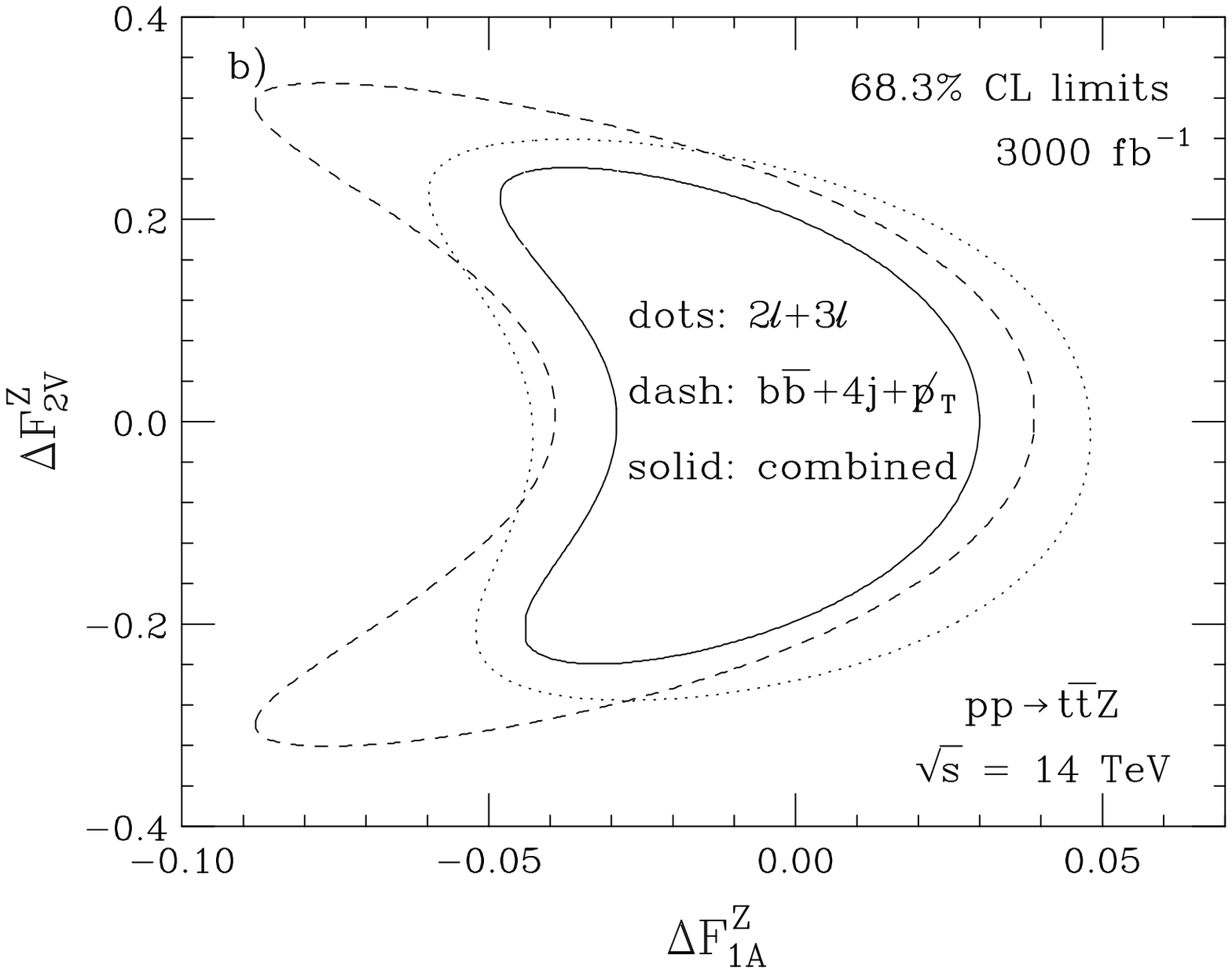} \\
\includegraphics[width=8.1cm]{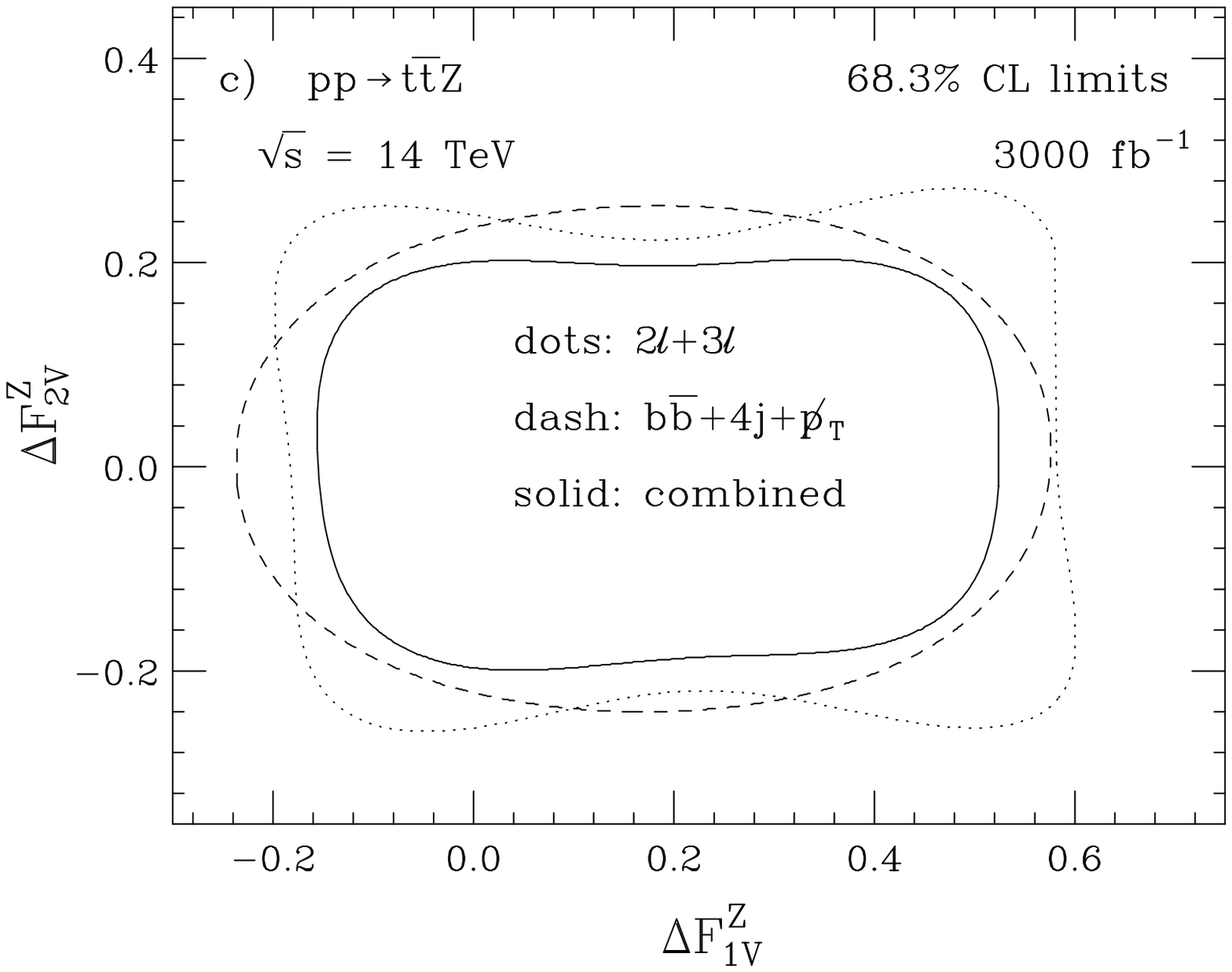} &
\includegraphics[width=8.1cm]{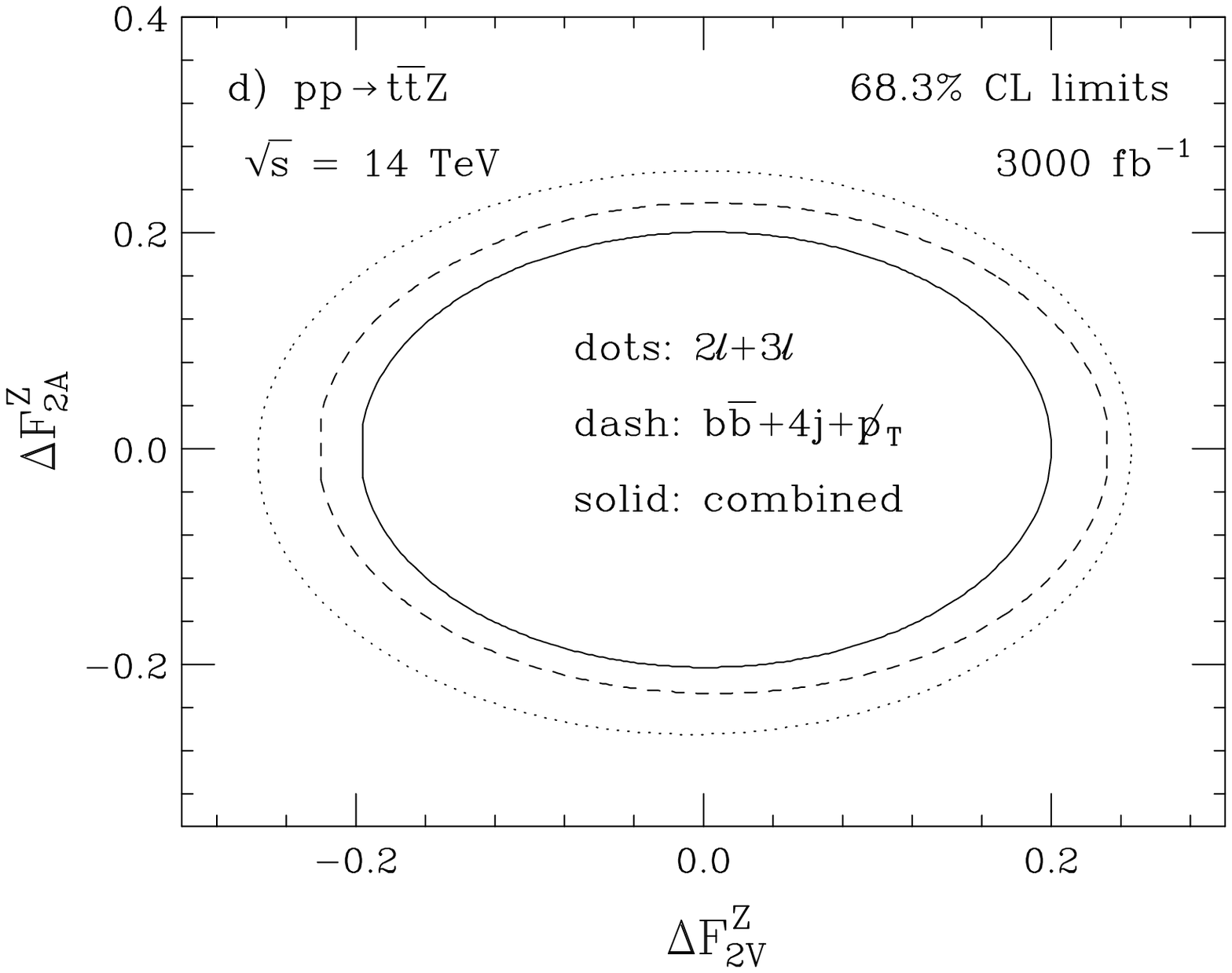} 
\end{tabular}
\caption[]{Projected $68.3\%$ CL bounds on anomalous $ttZ$ couplings
from the SLHC with an integrated luminosity of 3000~fb$^{-1}$, for:
(a) $\Delta F^Z_{1A}$ versus $\Delta F^Z_{1V}$, (b) $\Delta F^Z_{2V}$
versus $\Delta F^Z_{1A}$, (c) $\Delta F^Z_{2V}$ versus $\Delta
F^Z_{1V}$, and (d) $\Delta F^Z_{2A}$ versus $\Delta F^Z_{2V}$.  Shown
are the limits obtained from the $p\llap/_T b\bar{b}$+$4j$ (dashed)
and the dilepton and trilepton final states (dotted), and the combined
limits (solid).  To derive limits for the dilepton and trilepton final
states, we use the results of Ref.~\protect\cite{Baur:2004uw}.  In
each graph, only those couplings which are plotted against each other
are assumed to be different from their SM values.}
\label{fig:fig4}
\vspace{-7mm}  
\end{center}
\end{figure}
Our results are shown in Figs.~\ref{fig:fig3} and~\ref{fig:fig4}, and
in Table~\ref{tab:tab1}.  Fig.~\ref{fig:fig3} shows the correlations
between various anomalous $ttZ$ couplings for an integrated luminosity
of 300~fb$^{-1}$; Fig.~\ref{fig:fig4} displays the bounds one can hope
to achieve at the SLHC with 3000~fb$^{-1}$.  Shown are the results for
the $p\llap/_Tb\bar{b}$+$4j$ final state (dashed lines), the combined
limits from the dilepton and trilepton final states analyzed in
Ref.~\cite{Baur:2004uw} (dotted lines), and the limits resulting from
combining all final states (solid lines).  Including the
$p\llap/_Tb\bar{b}$+$4j$ final state in the extraction of bounds
improves the limits by $10-60\%$.  For an integrated luminosity of
300~fb$^{-1}$, it will be possible to measure the $ttZ$ axial vector
coupling with a precision of about $10\%$, and $F^Z_{2V,A}$ with a
precision of $40\%$.  At the SLHC, these bounds can be improved by
factors of about~1.6 ($F^Z_{2V,A}$) and~$2.3-3$ ($F^Z_{1A}$).  The
achievable bounds on $F^Z_{1V}$ are much weaker than those projected
for $F^Z_{1A}$: as mentioned in Sec.~\ref{sec:sec3}, the $p\llap/_T$
distributions for the SM and for $F^Z_{1V,A}=-F^{Z,SM}_{1V,A}$ are
almost degenerate.  As a result, an area centered at $\Delta
F^Z_{1V,A}=-2F^{Z,SM}_{1V,A}$ remains, which cannot be excluded, even
at the SLHC where one expects several thousand $t\bar{t}Z$ events.
For $F^Z_{1V}$, the two regions merge, resulting in rather poor
limits.  For $F^Z_{1A}$, the two regions are distinct.  Since the area
centered at $\Delta F^Z_{1A}=-2F^{Z,SM}_{1A}$ is incompatible with the
indirect limits on the $ttZ$ vector and axial vector couplings from
$Z$-pole data~\cite{Baur:2004uw}, we do not include this are in
Table~\ref{tab:tab1} or Figs.~\ref{fig:fig3} and~\ref{fig:fig4}.

While the bounds on $\Delta F^Z_{1A}$ improve by $20-60\%$ when the
$p\llap/_Tb\bar{b}$+$4j$ channel is included in the analysis, the gain
is limited to about $10\%$ for $\Delta F^Z_{1V}$.  The global limits
on the dimension-five couplings $F^Z_{2V,A}$ improve by about $20\%$.
However, if only $F^Z_{2V}$ and $F^Z_{2A}$ are varied, including the
$p\llap/_Tb\bar b$+$4j$ final state strengthens their bounds by about
$35\%$.  The relatively larger importance of the
$p\llap/_Tb\bar{b}$+$4j$ channel in this case can be understood by
recalling that most of the sensitivity to the dimension-five couplings
originates from high $p\llap/_T$ values where the background is less
important.  This allows one to take better advantage of the larger
$p\llap/_Tb\bar{b}$+$4j$ final state cross section.

The correlations between $F^Z_{2A}$ and $F^Z_{1A}$ ($F^Z_{2A}$ and
$F^Z_{1V}$) are similar to those for $F^Z_{2V}$ and $F^Z_{1A}$
($F^Z_{2V}$ and $F^Z_{1V}$), thus are not shown in
Figs.~\ref{fig:fig3} and~\ref{fig:fig4}.

\begin{table}
\begin{tabular}{|cccc|cccc|}
\hline
\multicolumn{4}{|c|}{300~fb$^{-1}$ (LHC) } & 
\multicolumn{4}{c|}{3000~fb$^{-1}$ (SLHC)} \\
\hline
\; coupling\; & $p\llap/_Tb\bar{b}$+$4j$\; & $2\ell$+$3\ell$ & \; combined\; &
\; coupling\; & $p\llap/_Tb\bar{b}$+$4j$\; & $2\ell$+$3\ell$ & \; combined\; \\
\hline 
$\Delta F^Z_{1V}$   & -- & $\begin{matrix} +0.84  \\[-4pt]
-0.43\end{matrix}$  & $\begin{matrix} +0.75  \\[-4pt]
-0.36\end{matrix}$  &
$\Delta F^Z_{1V}$   & -- & $\begin{matrix} +0.60  \\[-4pt]
-0.18\end{matrix}$  & $\begin{matrix} +0.54  \\[-4pt]
-0.16\end{matrix}$  \\
$\Delta F^Z_{1A}$   & $\begin{matrix} +0.12  \\[-4pt]
-\end{matrix}$      & $\begin{matrix} +0.16  \\[-4pt]
-0.13\end{matrix}$  & $\begin{matrix} +0.096 \\[-4pt]
-0.112\end{matrix}$ &
$\Delta F^Z_{1A}$   & $\begin{matrix} +0.047 \\[-4pt]
-\end{matrix}$      & $\begin{matrix} +0.049 \\[-4pt]
-0.060\end{matrix}$ & $\begin{matrix} +0.031 \\[-4pt]
-0.048\end{matrix}$ \\
$\Delta F^Z_{2V}$   & $\begin{matrix} +0.59  \\[-4pt]
-0.55\end{matrix}$  & $\begin{matrix} +0.47  \\[-4pt]
-0.47\end{matrix}$  & $\begin{matrix} +0.38  \\[-4pt]
-0.39\end{matrix}$  &
$\Delta F^Z_{2V}$   & $\begin{matrix} +0.34  \\[-4pt]
-0.32\end{matrix}$  & $\begin{matrix} +0.28  \\[-4pt]
-0.28\end{matrix}$  & $\begin{matrix} +0.25  \\[-4pt]
-0.24\end{matrix}$  \\
$\Delta F^Z_{2A}$   & $\begin{matrix} +0.57  \\[-4pt]
-0.58\end{matrix}$  & $\begin{matrix} +0.48  \\[-4pt]
-0.49\end{matrix}$  & $\begin{matrix} +0.40  \\[-4pt]
-0.40\end{matrix}$  &
$\Delta F^Z_{2A}$   & $\begin{matrix} +0.33  \\[-4pt]
-0.33\end{matrix}$  & $\begin{matrix} +0.28  \\[-4pt]
-0.29\end{matrix}$  & $\begin{matrix} +0.25  \\[-4pt]
-0.25\end{matrix}$  \\
\hline
\end{tabular}
\vspace{2mm}
\caption{Sensitivities achievable at $68.3\%$ CL for anomalous $ttZ$ 
couplings at the LHC and SLHC for integrated luminosities of
300~fb$^{-1}$, and 3000~fb$^{-1}$.  The limits shown represent the
maximum and minimum values obtained when taking into account the
correlations between any possible pair of anomalous couplings.
Results are for the $p\llap/_T b\bar{b}$+$4j$ final state, the
combined dilepton and trilepton final states analyzed in
Ref.~\protect\cite{Baur:2004uw} (labeled $2\ell+3\ell$), and for all
channels combined.  The cuts imposed are described in
Sec.~\ref{sec:sec2b}.}
\label{tab:tab1}
\end{table}

The $ttZ$ couplings are indirectly constrained by precision $Z$-pole
data collected at LEP and SLC.  Vector
and axial vector couplings are bound by $Z$-boson data to within a few
percent of their SM values if one assumes that no other sources of new
physics contribute.  In contrast, the limits obtained for $F^Z_{2V}$
are much weaker, $|F^Z_{2V}|\lesssim {\cal O}(0.2)$, and depend on the
value of the anomalous magnetic moment of the top
quark~\cite{Eboli:1997kd}.  The effect of $F^Z_{2A}$ on LEP/SLC
observables has not yet been studied.  Thus, $t\bar{t}Z$ production at
the LHC will provide valuable information on the dimension-five
couplings.  Since the LEP/SLC constraints arise from one-loop corrections
which diverge for non-standard $ttZ$ couplings, they are
cutoff-dependent.  When taking into account the
$p\llap/_Tb\bar{b}$+$4j$ final state, the achievable sensitivity on
$F^Z_{1A}$ at the SLHC begins to approach that of the indirect bounds
from $Z$-pole data.  For the $ttZ$ vector coupling, it will be impossible
to match that precision at the LHC, even for an integrated luminosity
of 3000~fb$^{-1}$ and including the $p\llap/_Tb\bar b$+$4j$ final
state in the analysis.

The $ttZ$ couplings can also be tested in $e^+e^-\to t\bar{t}$.
However, as mentioned in Sec.~\ref{sec:sec1}, the process
$e^+e^-\to\gamma^*,\,Z^*\to t\bar{t}$ is sensitive to both $tt\gamma$
and $ttZ$ couplings simultaneously.  If only one coupling at a time is
allowed to deviate from its SM value, a linear $e^+e^-$ collider
operating at $\sqrt{s}=500$~GeV with an integrated luminosity of
$100-200$~fb$^{-1}$ would be able to probe {\sl all} $ttZ$ couplings
with a precision of $1-5\%$~\cite{Abe:2001nq}.  With the possible
exception of $F^Z_{1A}$, a linear collider will thus be able to
significantly improve the sensitivity limits expected from the LHC,
even when the $p\llap/_Tb\bar{b}$+$4j$ channel is included and
assuming 3000~fb$^{-1}$ from the SLHC.  It should be noted, however,
that this picture could change once cancellations between different
non-standard $ttZ$ couplings, and between $tt\gamma$ and $ttZ$
couplings, are allowed.  While beam polarization at an $e^+e^-$ collider
would provide a very powerful tool to disentangle the effects of
different couplings, unfortunately no realistic studies on 
the simultaneous measurement of couplings in $e^+e^-\to
t\bar{t}$ have been performed so far.

%------------------------------------------------------------------------
\subsection{Constraints on Little Higgs parameter space}
%------------------------------------------------------------------------

As noted in Sec.~\ref{sec:sec1}, the $ttZ$ couplings may deviate
substantially from their SM values in Little Higgs (LH) models.  Here
we explore how their measurement at the LHC may constrain parameter
space in the $SU(5)/SO(5)$ Littlest Higgs model with
T-parity~\cite{Cheng:2003ju,Cheng:2004yc,Low:2004xc}.  In this model,
the $ttZ$ vector and axial vector couplings are modified by mixing of
the left-handed top quark and a heavy top quark partner, $T$.  One
finds
\begin{equation}\label{eq:LH}
\Delta F^Z_{1V} = -\Delta F^Z_{1A} = \frac{\lambda_T^2v^2}{2m_T^2}
                                     \,F^{Z,SM}_{1A} \, ,
\end{equation}
where $\lambda_T$ is the $tTh$ coupling ($h$ is the Higgs boson),
$v\approx 246$~GeV is the SM Higgs vacuum expectation value, and $m_T$
is the $T$ quark mass.  Eq.~(\ref{eq:LH}) and the log likelihood
function can be used to derive lower $68.3\%$ CL limits for
$m_T/\lambda_T$:
\begin{eqnarray}
\frac{m_T}{\lambda_T} & \geq &  600~{\rm GeV}\qquad {\rm for~300~fb^{-1},} \\
\frac{m_T}{\lambda_T} & \geq & 1000~{\rm GeV}\qquad {\rm for~3000~fb^{-1}.}
\end{eqnarray}
For comparison, current EW precision data require
$m_T/\lambda_T>650$~GeV~\cite{Hubisz:2005tx}.  We thus expect the SLHC
will be able to improve this bound, while the LHC should be able to
discover a $T$ quark with a mass of $m_T\leq 2$~TeV with 300~fb$^{-1}$
of data~\cite{Azuelos:2004dm}.  If a $T$-quark candidate were found, a
measurement of $F^Z_{1A}$ would be valuable in helping to pin down
$\lambda_T$.

In LH models without T-parity, anomalous $ttZ$ couplings may receive
additional contributions from mixing of the $W$ and $Z$ bosons with a
heavy $SU(2)$ triplet of vector bosons, $W^\pm_H$ and $W^3_H$, which
are characteristic for LH models~\cite{Berger:2005ht}.  However,
constraints from precision EW data severely restrict these additional
contributions.

%------------------------------------------------------------------------
\section{Summary and Conclusions}
\label{sec:sec4}
%------------------------------------------------------------------------

Currently, little is known about top quark couplings to the $Z$ boson.
There are no direct measurements of these couplings; indirect
measurements, using LEP and SLC data, tightly constrain only the $ttZ$ vector
and axial vector couplings.  The $ttZ$ couplings could be measured
directly in $e^+e^-\to t\bar{t}$ at a future $e^+e^-$ linear collider.
However, such a machine is at least a decade away.  In addition, the
process $e^+e^-\to t\bar{t}$ is simultaneously sensitive to $tt\gamma$
and $ttZ$ couplings, and significant cancellations between various
couplings may occur.

In this paper, we considered $t\bar{t}Z$ production with
$Z\to\bar\nu\nu$ and $t\bar{t}\to b\bar{b}$+$4j$ at the LHC as a tool
to measure $ttZ$ couplings.  At the Tevatron, the $t\bar{t}Z$ cross
section is too small to be observable.  When the $Z$ boson decays
leptonically, the small $Z\to\ell^+\ell^-$ branching ratio is one of
the main factors which limits the achievable sensitivity to anomalous
$ttZ$ couplings.  The larger $Z\to\bar\nu\nu$ branching ratio
(relative to $Z\to\ell^+\ell^-$) effectively triples the number of
signal events and thus has the potential to significantly improve the
sensitivity to non-standard couplings.

We calculated the signal cross sections taking into account all top
quark-resonant Feynman diagrams.  All relevant background processes
were included in estimating limits on the couplings.  Once
$t\bar{t}Z$ selection cuts are imposed, the background drops
significantly faster with missing transverse momentum than the signal,
and for $p\llap/_T>380$~GeV the signal dominates.  The largest
background source is $t\bar{t}jj$ production, where one of the top
quarks decays semi-leptonically and the charged lepton is lost.  In
all our calculations we assumed that both $b$ quarks are tagged, to
bring the backgrounds to a controllable level.

Our analysis reveals that the achievable sensitivity limits utilizing
final states where the $Z$ decays leptonically~\cite{Baur:2004uw} can
be improved by $10-60\%$ when the $p\llap/_Tb\bar{b}$+$4j$ mode is
taken into account.  The improvement is particularly pronounced for
the $ttZ$ axial vector coupling $F^Z_{1A}$ which can be measured with
a precision of $3-5\%$ at the luminosity-upgraded LHC
(3000~fb$^{-1}$).  Measuring $F^Z_{1A}$ with such precision will make
it possible constrain the parameter space in Little Higgs models which
predict deviations of the $ttZ$ vector and axial vector couplings of
up to $10\%$.

%------------------------------------------------------------------------

\begin{acknowledgments}
We would like to thank J.~Parsons, G.~Watts and J.~Womersley for
useful discussions.  One of us (U.B.) would like to thank the Fermilab
Theory Group, where part of this work was carried out, for its
generous hospitality.  This research was supported in part by the
National Science Foundation under grant No.~PHY-0139953 and the
Department of Energy under grant DE-FG02-91ER40685.  Fermilab is
operated by Universities Research Association Inc. under Contract
No. DE-AC02-76CH03000 with the U.S. Department of Energy.
\end{acknowledgments}

%------------------------------------------------------------------------

%%%%%%%%%%%%%%%%%%%%%%%%%%%%%%%%%%% References %%%%%%%%%%%%%%%%%%%%%%%%%%%%%%%%%%%

%\bibliographystyle{plain}

\end{document}